\documentclass[conference]{IEEEtran}
\pdfoutput=1
\usepackage{cite}
\usepackage{amsmath,amssymb,amsfonts}
\usepackage{algorithm}
\usepackage{algorithmic}
\usepackage{graphicx}
\usepackage{times}
\usepackage{epsfig}

\usepackage{textcomp}
\usepackage{xcolor}
\def\BibTeX{{\rm B\kern-.05em{\sc i\kern-.025em b}\kern-.08em
    T\kern-.1667em\lower4.7ex\hbox{E}\kern-.125emX}}

\usepackage{CJKutf8}

\usepackage{threeparttable}
\usepackage{multirow}
\usepackage{enumitem}
\usepackage{caption}
\usepackage{subfloat}
\usepackage{url}
\usepackage{xpinyin}

\begin{document}

\title{
Exploring Text-based Realistic Building Facades Editing Applicaiton
}
\author{
	\IEEEauthorblockN{
		Jing~Wang, and
        Xin~Zhang
	}

	\IEEEauthorblockA{
		\textit{HUST, China}
		}
}

\twocolumn[{

\renewcommand\twocolumn[1][]{#1}%
\maketitle

\begin{center}
    \centering
    \includegraphics[width=0.95\textwidth]{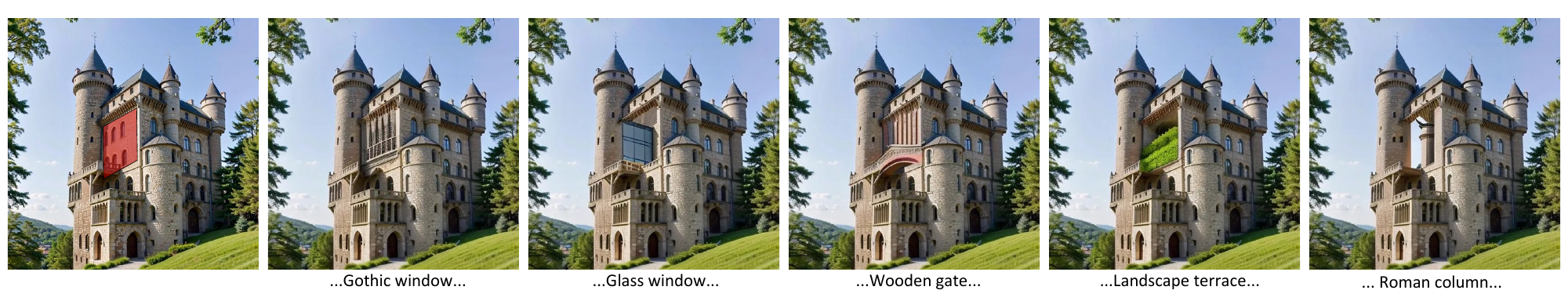}
    \captionsetup{type=figure}\caption{
Through the integration of the powerful editing capabilities of diffusion models and user-drawn masks, we have achieved precise editing of architectural facades. This innovative approach opens up new possibilities in the field of architectural design, allowing designers to customize the appearance of buildings more intuitively and efficiently. In this editing process, users first provide textual descriptions of the architectural facade editing effects they wish to achieve. Subsequently, users specify the editing areas by drawing masks, such as for windows, doors, walls, etc.
    }
    \label{fig:teaser}
\end{center}
}]

\begin{figure*}[t]
    \centering

    \includegraphics[width=0.84\textwidth]{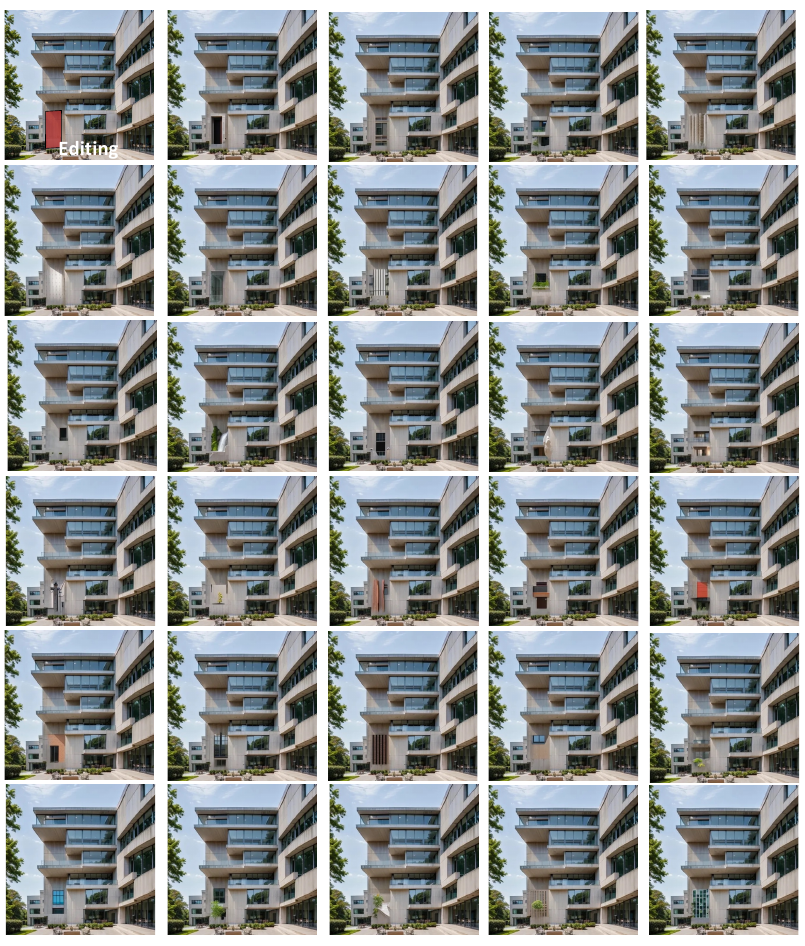}
    \caption{We can control the editing areas and achieve local modifications of building facades via user-drawn masks and prompts}

\end{figure*}  

\begin{figure*}[t]
    \centering

    \includegraphics[width=0.83\textwidth]{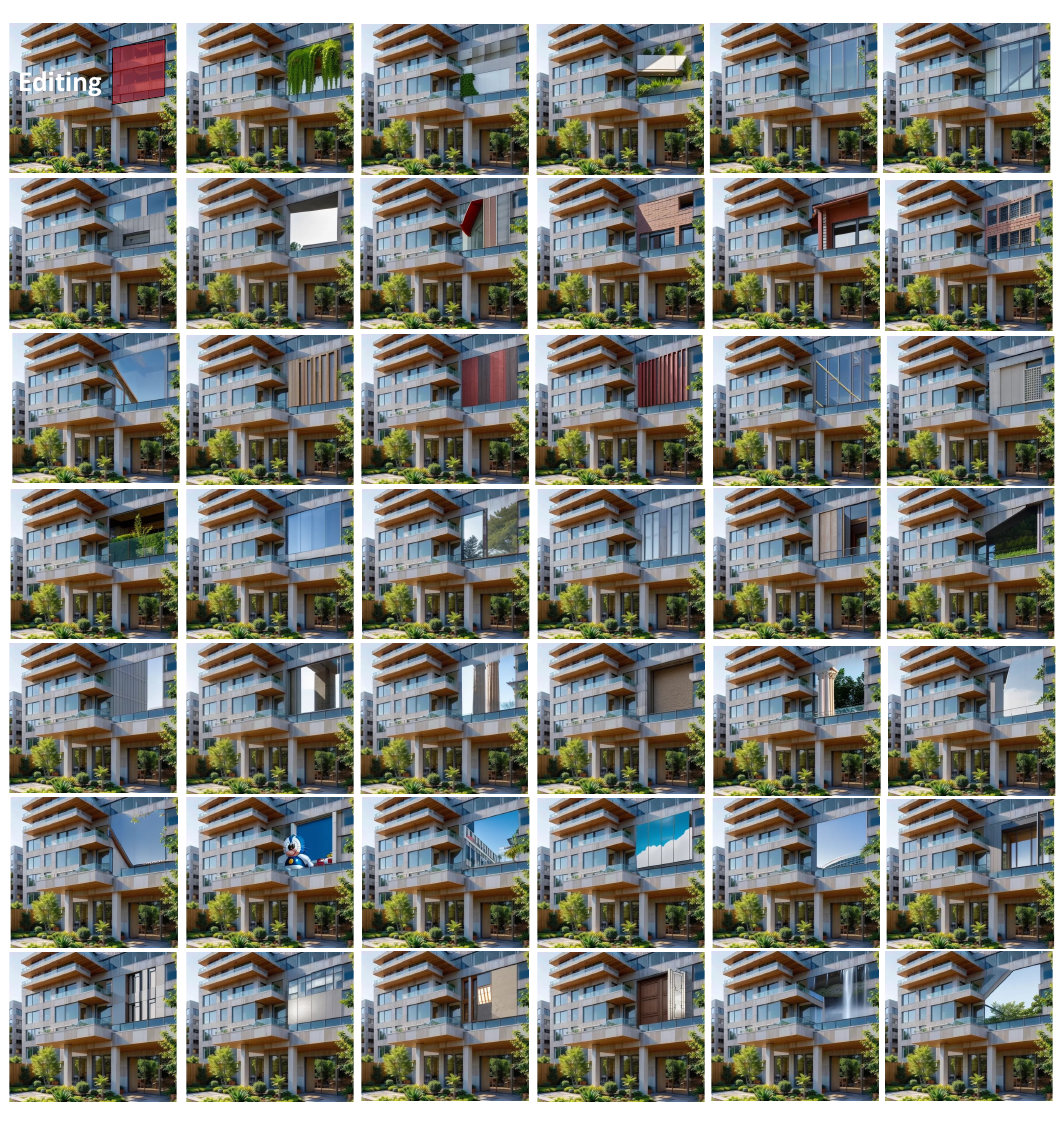}
    \caption{Diffusion models possess powerful generative capabilities, allowing them to realistically synthesize details and structures of architectural facades. By combining with user-drawn masks, we can precisely control the editing areas and achieve local modifications of building appearances. This method not only enables rapid generation of edited images but also preserves the realism and quality of the images.}
   
\end{figure*}  

\begin{abstract}
This paper explores the utilization of diffusion models and textual guidance for achieving localized editing of building facades, addressing the escalating demand for sophisticated editing methodologies in architectural design and urban planning. Leveraging the robust generative capabilities of diffusion models, this study presents a promising avenue for realistically synthesizing and modifying architectural facades. Through iterative diffusion and text descriptions, these models adeptly capture both the intricate global and local structures inherent in architectural facades, thus effectively navigating the complexity of such designs. Additionally, the paper examines the expansive potential of diffusion models in various facets, including the generation of novel facade designs, the enhancement of existing facades, and the realization of personalized customization. Despite their promise, diffusion models encounter obstacles such as computational resource constraints and data imbalances. To address these challenges, the study introduces the innovative Blended Latent Diffusion method for architectural facade editing, accompanied by a comprehensive visual analysis of its viability and efficacy. Through these endeavors, we aims to propel forward the field of architectural facade editing, contributing to its advancement and practical application.
\end{abstract}

\section{Introduction}
\label{sec:intro}
In recent years, with the rapid development of  artificial intelligence 2D/3D technology \cite{Rae2021ScalingLM,li2023towards,Thoppilan2022LaMDALM, Brown2020LanguageMA} and deep learning technologies~\cite{rombach2022high,he2016deep,he2017mask},there has been a growing demand for facade editing techniques in the fields of architectural design and urban planning~\cite{groat2013architectural,garlan1994exploiting,demirbacs2003focus,caetano2020computational,aliakseyeu2006computer,akin1996frames,lomas2007architectural,thalfeldt2013facade,lim2012building}. The facade of a building, as its external interface, not only directly impacts the aesthetic appearance of the building but also bears many functional and environmental responsibilities. Therefore, effectively synthesizing and editing building facades has become a critical aspect of the architectural design process.

Deep learning technology has made breakthroughs in the field of image synthesis and editing~\cite{li2023layerdiffusion,brooks2023instructpix2pix,parmar2023zero,mokady2023null,chang2023muse,ruiz2023dreambooth,chefer2023attend,kumari2023multi,karras2019style}. Among them, diffusion models have gained considerable attention as an emerging deep generative model. Compared to traditional Generative Adversarial Networks (GANs)~\cite{goodfellow2020generative,creswell2018generative,liu2016coupled}, diffusion models demonstrate better stability and sample quality in image generation, making them a potentially powerful application for architectural facade editing.

Facade editing based on diffusion models involves iteratively diffusing and updating the information of an image to gradually generate realistic images. Unlike traditional generative models, diffusion models can gradually generate images by iterating multiple times on each pixel, allowing them to better capture the global and local structures of images, especially suitable for editing the complex structures and details of architectural facades.

In the field of architectural facade editing~\cite{shalunts2011architectural,lim2012buildingfacade,moghtadernejad2019facade,hosseini2019morphologicaldfacade,thalfeldt2013facade,leskovar2011approach}, diffusion models hold broad application prospects~\cite{li2023sketch,li2024generating}. Firstly, they can be used to generate new designs for building facades, providing architects with more diverse design options. Secondly, diffusion models can also be used to edit the appearance details of existing building facades, such as adjusting colors, materials, textures, etc., thereby achieving fine control over the building's appearance. Additionally, diffusion models can perform local editing based on user needs, such as adding or removing building elements at specific locations, enabling personalized customization of the building's appearance. Compared to traditional methods of architectural facade editing, diffusion model-based editing methods~\cite{li2023sketch,li2024generating} have many advantages. Firstly, their powerful generative capability can produce more realistic and detailed architectural facade images, enhancing the realism and visual quality of the editing results. Secondly, some diffusion model-based methods can perform fine adjustments to architectural facades while maintaining the integrity of the image structure. Furthermore, diffusion models can also incorporate natural language descriptions, achieving semantic-level editing of architectural facades, thereby enhancing the flexibility and intelligence of editing.

However, despite the enormous potential of diffusion models in architectural facade editing, they still face many challenges in practical applications. Firstly, due to the complexity and diversity of architectural facades, effectively representing and capturing their features and structures becomes a key issue. Secondly, the training and inference processes of diffusion models often require substantial computational resources and time, limiting their scalability and practicality in real-world applications. Additionally, when dealing with large-scale architectural facade data, diffusion models often encounter problems such as data imbalance and sample scarcity, which also constrain their effectiveness and performance in practical applications.

As shown in Fig. 1, this paper explores the latest image editing techniques, applying the Blended Latent Diffusion method to architectural facade editing, and provides a detailed visual analysis of its feasibility and generation effects in architectural design. It aims to provide new editing tools and methods for the field of architectural design and planning and proposes corresponding solutions and improvement methods. Through this research, we aim to provide a deeper understanding and exploration for the further development and application of architectural facade editing.

\section{Related Method}
\label{sec:RW}
In this section, we introduce a image editing method called Blended Latent Diffusion~\cite{avrahami2022blended,avrahami2023blended}, which guides image generation and editing through text descriptions. This method combines the capabilities of diffusion models~\cite{rombach2022high} and visual-language models, featuring the following key steps and characteristics:

\paragraph{Local Editing Scenario} Unlike most text-guided methods that focus on generating images from scratch or globally modifying existing ones, Blended Latent Diffusion emphasizes local editing scenarios. This means artists only need to modify part of an image while retaining the rest.

\paragraph{Diffusion Models} The method is based on diffusion models, known for their excellent performance in generation, editing, and other tasks. However, the iterative diffusion process at the pixel level results in longer inference times. Therefore, recent research proposes conducting diffusion in lower-dimensional and higher-level semantic latent spaces to achieve competitive performance in shorter training and inference times.

\paragraph{Text Guidance} Blended Latent Diffusion utilizes text prompts to guide image generation and editing. Users can add new objects, modify existing ones, or inject textual content into images by providing text prompts, achieving intuitive and creative editing effects.

\paragraph{Background Optimization}The method supports whether to optimize the background during generation, allowing users to decide on this additional step as needed. Even considering background optimization, Blended Latent Diffusion outperforms baseline methods in standard batch inference scenarios, with inference speeds ten times faster than Blended Diffusion and Local CLIP-guided diffusion.

\paragraph{Ranking Effects} By ranking generated results using the CLIP model, the study found that the top 20$\%$ of results are generally better than the bottom 20$\%$, although not every image strictly surpasses another.

Blended Latent Diffusion~\cite{avrahami2023blended} combines diffusion models and text-guided techniques to achieve the goal of generating and editing images in local editing scenarios, providing users with an intuitive and efficient image processing tool. Users simply describe the architectural facade editing effects they want to achieve, such as adjusting appearance details or modifying materials and colors, and the system generates corresponding editing results based on the provided text prompts. This text-based editing approach streamlines the designer's workflow, significantly speeding up the facade design process. The following sections will detail how this algorithm is applied to architectural facade design.

\section{Facade Editing}

The application of the Blended Latent Diffusion~\cite{avrahami2022blended} method in the field of architectural facade design provides designers and architects with an innovative way to achieve local editing of facades. By combining text guidance and image generation techniques, this method enables users to make design modifications quickly and intuitively, explore different design concepts, and achieve personalized customization of architectural appearances.

Firstly, designers can input the architectural facade images that need to be edited. These could be existing architectural sketches, actual building photos, or digital modeling images. Using the Blended Latent Diffusion method, designers can perform local editing on specific areas such as windows, doors, decorative elements, while retaining the original architectural structure. This flexibility in local editing allows designers to quickly experiment with different design concepts and explore various possibilities.

Secondly, users can guide the editing process by providing text prompts. These prompts describe the desired design effects, such as "adding a large glass window on the left side of the architectural facade" or "adding a protruding decorative element at the top of the building". This text-guided approach makes the editing process more intuitive and easy to understand, while also providing users with more creative inspiration.

During the editing process, users can mark the areas to be edited to specify the editing scope. This marking can be done using simple drawing tools or editing software to ensure the accuracy and precision of the editing. By combining text prompts and marking of editing areas, the Blended Latent Diffusion method can generate edited architectural facade images according to user needs.

After generating the results, users can adjust and optimize the edited architectural facade images. This may involve fine-tuning aspects such as color, texture, shape, etc., to ensure that the editing effects meet expectations. Designers can further modify the generated images as needed to meet design requirements and aesthetic standards.

As shown in Fig. 2 and Fig. 3, through the application of the Blended Latent Diffusion method in the field of architectural facade design, designers and architects can conduct local facade editing more quickly and efficiently, achieving rapid iteration and validation of design ideas. This method not only helps designers save time and effort but also provides them with more design possibilities and creative space, thereby driving innovation and development in the field of architectural design.

In summary, the Blended Latent Diffusion method brings new possibilities and opportunities to the field of architectural facade design, providing designers with an intuitive and efficient tool for local facade editing and promoting innovation and personalized customization in architectural design. By combining text guidance and image generation techniques, this method injects new vitality and creativity into the field of architectural design, opening up more possibilities and imagination for future architectural design.

\section{Conclusion}
In conclusion, this paper focus on the Blended Latent Diffusion method demonstrates its potential for enhancing architectural facade editing, providing a detailed visual analysis of its feasibility and generation effects. By proposing solutions and improvement methods, the study contributes to the advancement of architectural design and planning tools.  the application prospects of diffusion models in architectural facade editing are discussed comprehensively, ranging from generating new designs to fine-tuning existing facade details and performing localized edits based on user preferences. Despite their advantages, diffusion models face challenges such as effectively representing complex facade features and requiring significant computational resources for training and inference. Overall, this paper underscores the importance of exploring innovative image editing techniques like the Blended Latent Diffusion method in addressing the evolving needs of architectural facade editing, ultimately aiming to facilitate deeper understanding and advancement in this critical area of architectural design.

\bibliographystyle{IEEEtran}
\bibliography{IEEEabrv,custom}

\end{document}